\documentclass[epj]{svjour}
\usepackage{graphicx}
\usepackage{bm}
\usepackage{amsmath}
\usepackage{amssymb}
\begin{document}
\title{Chiral dynamics and $s$-wave exotic hadrons}
% \subtitle{Do you have a subtitle?\\ If so, write it here}
\author{Tetsuo~Hyodo\inst{1} \and Daisuke~Jido\inst{1}
\and Atsushi~Hosaka\inst{2}% etc
% \thanks is optional - remove next line if not needed
% \thanks{\emph{Present address:} Insert the address here if needed}%
}                     % Do not remove
%
% \offprints{}          % Insert a name or remove this line
%
\institute{Yukawa Institute for Theoretical Physics, 
Kyoto University, Kyoto 606--8502, Japan 
\and Research Center for Nuclear Physics (RCNP),
Ibaraki, Osaka 567-0047, Japan.}
\date{Received: date / Revised version: date}
% The correct dates will be entered by Springer
%
\abstract{
    Based on chiral dynamics, existence of exotic hadrons is discussed in 
    the SU(3) symmetric limit. The low energy $s$-wave interaction of the 
    Nambu-Goldstone boson with a hadron is known to be determined 
    model-independently, which has been used to generate some hadron 
    resonances in nonexotic channels. We show that this interaction in any 
    exotic channels is not strong enough to generate a bound state.
 % end of PACS codes
} %end of abstract
\maketitle

Experimentally, more than hundred of hadrons have been discovered so far, 
whose properties are summarized by the Particle Data Group 
(PDG)~\cite{Yao:2006px}. Most of the hadrons can be, in principle, described
in terms of $\bar{q}q$ or $qqq$, and the only state with exotic flavor 
quantum numbers in PDG is the $S=+1$ baryon 
$\Theta^{+}$~\cite{Nakano:2003qx}. Thus, exotic hadrons are indeed 
``exotic'' from an experimental point of view, while there is no clear 
theoretical explanation of the non-observation (or non-existence) of the 
exotic hadrons. This is certainly a non-trivial issue to be explained 
theoretically, irrespective to the existence of the $\Theta^+$.

Here we report our recent work on this 
issue~\cite{Hyodo:2006yk,Hyodo:2006kg}, in which we have studied the 
existence of the exotic hadrons in $s$-wave scattering of a hadron and the 
Nambu-Goldstone (NG) boson. We utilize the theoretical framework based on the
$S$-matrix theory. In 60's, the framework was used to describe hadron 
resonances such as $\Lambda(1405)$~\cite{Dalitz:1960du,PR155.1649} with the 
effective vector meson exchange interaction. Recently, the interaction has 
been founded by chiral 
symmetry~\cite{Kaiser:1995eg,Oset:1998it,Oller:2000fj,Lutz:2001yb}, leading 
to a successful description of the hadron resonances in chiral unitary 
approaches~\cite{Jido:2003cb,Hyodo:2003jw,Magas:2005vu}.

In recent applications of the chiral unitary approach, it was shown that 
some resonances obtained in the coupled channel dynamics with SU(3) breaking 
became bound states of a single channel in the flavor SU(3) 
limit~\cite{Jido:2003cb,Garcia-Recio:2003ks,Kolomeitsev:2003kt,Sarkar:2004jh,Lutz:2003jw,Kolomeitsev:2003ac}.
Therefore, we expect that the origin of the physical resonances may be 
clarified by studying the bound states in the SU(3) limit. We focus on the 
$s$-wave scatterings, since the low energy interaction of the NG boson with 
any hadrons in $s$-wave are uniquely determined by chiral symmetry.
In this framework, we examine the possibility to generate the exotic hadron
as a bound state of the NG boson and a hadron.

% WT interaction

\begin{figure}[tbp]
\centerline{\includegraphics[width=0.5\textwidth]{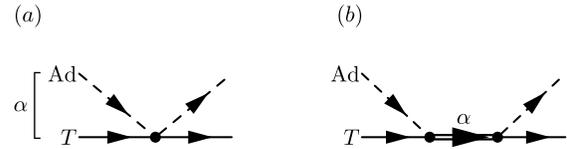}}
\caption{(a) : Notation of the representations $\alpha$, Ad and $T$ for 
the WT term. (b) : The bound state pole diagram after unitarization of 
the amplitude.}
\label{fig:Rep}
\end{figure}

The low energy $s$-wave interaction of the NG boson~(Ad) with a target 
hadron~($T$) is model-independently given by the Weinberg-Tomozawa 
theorem~\cite{Weinberg:1966kf,Tomozawa:1966jm} as
\begin{equation}
    V_{\alpha }
    =-\frac{ \omega}{2f^2}C_{\alpha,T}  , 
    \label{eq:WTint}
\end{equation}
with the decay constant $(f)$ and energy $(\omega)$ of the NG boson.
The group theoretical factor $C_{\alpha,T}$ is determined by specifying 
the flavor representations of the target $T$ and the scattering system 
$\alpha\in T\otimes $Ad (see Fig.~\ref{fig:Rep}):
\begin{equation}
    C_{\alpha,T}
    = - \left\langle 2{\bm F}_{T} \cdot {\bm F}_{\rm Ad}  
     \right\rangle_{\alpha} 
    =  C_2(T)-C_2(\alpha)+3 ,
    \label{eq:WTintfinal}
\end{equation}
where $C_2(R)$ is the quadratic Casimir of SU(3) for the representation $R$,
and we use $C_{2}({\rm Ad}) = 3$ for the adjoint representation of the NG 
boson.  

%--figure---------------------------------
%--figure---------------------------------

For the target hadron with an arbitrary SU(3) representation $T=[p,q]$, 
possible representations $\alpha$ for the scattering channels are obtained as
\begin{eqnarray}
  \lefteqn{
    [p,q]\otimes [1,1] =    [p+1,q+1]   
    } && \nonumber \\   && 
    \oplus [p+2,q-1]
    \oplus [p-1,q+2]
    \oplus [p,q]
    \oplus [p,q] \nonumber \\ &&
    \oplus [p+1,q-2]
    \oplus [p-2,q+1] 
     \oplus [p-1,q-1] ,
    \nonumber
\end{eqnarray}
where the labels of representations $[a,b]$ should satisfy $a, b \geq 0$, and
one of the two $[p,q]$ representations on the right hand side should satisfy
$p\geq 1$ and the other $q\geq 1$. Using Eq.~\eqref{eq:WTintfinal}, we 
evaluate the coupling strengths $C_{\alpha,T}$ for the possible 
representations of the channel~$\alpha$ as shown in Table~\ref{tbl:ECtable}.

% For tables use
\begin{table}
\caption{Properties of the WT interaction in the channel~$\alpha$ of the 
    NG boson scattering on the target hadron with the $T=[p,q]$ 
    representation. The coupling strengths of the WT term is denoted as 
    $C_{\alpha,T}$, $\Delta E$ is the differences of the exoticness $E$ 
    between the channel $\alpha$ and the target hadron $T$, and 
    $C_{\alpha,T}(N_c)$ denotes the coupling strengths for arbitrary $N_c$.}
\label{tbl:ECtable}       % Give a unique label
% For LaTeX tables use
\centering
\begin{tabular}{|c|ccc|}
    \hline
	$\alpha$  
	& $C_{\alpha,T}$ 
	& $\Delta E$ 
	& $C_{\alpha,T}(N_c)$  \\
        \hline
        $[p+1,q+1]$ 
	& $-p-q$ 
	& 1 or 0 
	& $\frac{3-N_c}{2}-p-q$  \\
        $[p+2,q-1]$ 
	& $1-p$ 
	& 1 or 0
	& $1-p$  \\
        $[p-1,q+2]$ 
	& $1-q$ 
	& 1 or 0 
	& $\frac{5-N_c}{2}-q$ \\
        $[p,q]$ 
	& $3$  
	& 0 
	& 3 \\
        $[p+1,q-2]$ 
	& $3+q$ 
	& 0 or $-1$ 
	& $\frac{3+N_c}{2}+q$  \\
        $[p-2,q+1]$ 
	& $3+p$  
	& 0 or $-1$ 
	& $3+p$ \\
        $[p-1,q-1]$ 
	& $4+p+q$  
	& 0 or $-1$ 
	& $\frac{5+N_c}{2}+p+q$ \\
	\hline
    \end{tabular}
% Or use
% \vspace*{5cm}  % with the correct table height
\end{table}

In order to specify the exotic channels, we define the exoticness quantum 
number 
$E$~\cite{Kopeliovich:1990ez,Diakonov:2003ei,Kopeliovich:2003he,Jenkins:2004tm}
as the number of valence quark-antiquark pairs to construct the 
given flavor multiplet $[p,q]$ with the baryon number $B$ carried by the $u$,
$d$, and $s$ quarks. For $B>0$ the exoticness $E$ is given by
\begin{equation}
    E=\epsilon\theta(\epsilon)+\nu\theta(\nu) ,
    \nonumber
\end{equation}
where we define $\epsilon \equiv (p+2q)/3-B$, $\nu\equiv (p-q)/3-B$ and 
$\theta(x)$ is the step function. For each $\alpha$, we evaluate the 
difference of the exoticness $\Delta E$ between the target $T$ and $\alpha$, 
as shown in Table~\ref{tbl:ECtable}. Taking $p,q \geq 0$ into account, we 
find that most of exotic channels are repulsive $(C_{\alpha,T}<0)$ and that 
the attractive interaction $(C_{\alpha,T}>0)$ is realized with the universal 
strength
\begin{equation}
    C_{\text{exotic}}=1  . \label{eq:Exoticattraction}
\end{equation}
By looking at the construction in detail, we can further specify the 
attractive channels as $\alpha=[p-1,2]$ for $T=[p,0]$ and 
$p\geq 3B$~\cite{Hyodo:2006yk,Hyodo:2006kg}.

Next we study the scattering problem with the WT interaction $V_{\alpha}$ of 
Eq.~\eqref{eq:WTint} in order to examine whether the attractive 
interaction~\eqref{eq:Exoticattraction} can accommodate an exotic bound 
state. We utilize the N/D method~\cite{Oller:2000fj} to construct the 
scattering amplitude which satisfies the elastic unitarity condition. The 
scattering amplitude of the NG boson and the target hadron in the channel 
$\alpha$ is given by
\begin{equation}
    t_{\alpha}(\sqrt{s})=
    \frac{1}{1-V_{\alpha}(\sqrt{s})G(\sqrt{s})}V_{\alpha}(\sqrt{s}) ,
    \nonumber
\end{equation}
as a function of the center-of-mass energy $\sqrt{s}$. Here 
$G(\sqrt{s})$ is
given by the once-subtracted dispersion integral
\begin{equation}
    G(\sqrt{s})
    =-\tilde{a}(s_0)
    -\frac{1}{2\pi}
    \int_{s^{+}}^{\infty}ds^{\prime}
    \left(
    \frac{\rho(s^{\prime})}{s^{\prime}-s}
    -\frac{\rho(s^{\prime})}{s^{\prime}-s_0}
    \right) ,
    \nonumber
\end{equation}  
with $\rho(s)=2M_{T}\sqrt{(s-s^+)(s-s^-)}/(8\pi s)$ and 
$s^{\pm}=(m\pm M_{T})^2$. 

The subtraction constant $\tilde{a}(s_0)$ should be in principle determined 
so as to reproduce some experimental observables. Here we would like to 
discuss a prescription which can be applicable to the cases without 
experimental information. For this purpose, we adopt the prescription given 
in Refs.~\cite{Igi:1998gn,Lutz:2001yb}:
\begin{equation}
    G(M_{T})=0  ,
    \label{eq:regucond} 
\end{equation}
which is equivalent to $t_{\alpha}(\sqrt s)=V_{\alpha}(\sqrt s)$ at 
$\sqrt s =M_{T}$. This allows us to match the amplitude with that of the 
u-channel ressumation at this energy. In addition, we show that this 
prescription provides a ``natural size'' of the subtraction 
parameter~\cite{Hyodo:2006kg}, with which the experimental observables in 
the $S=-1$ meson-baryon channel are well reproduced~\cite{Oller:2000fj}.

Since the WT interaction is the leading order term of the chiral 
perturbation theory, the condition~\eqref{eq:regucond} means that the full 
amplitude becomes that of the chiral perturbation theory at $\sqrt s =M_{T}$.
It is worth noting that $t_{\alpha}(\sqrt s)=V_{\alpha}(\sqrt s)$ can be 
achieved only in the region where $G(\sqrt{s})$ is real, since the 
subtraction constant $\tilde{a}(s_0)$ is a real number. Therefore, in order 
to match the full amplitude to the chiral perturbation theory, we need
\begin{equation}
    G(\sqrt{s})=0 \quad \text{within} \quad M_T-m \leq \sqrt{s} \leq M_T+m
    . \nonumber
\end{equation}
If this condition is not satisfied, the unitarized amplitude
$t_{\alpha}(\sqrt s)$ does not coincide with the tree level amplitude 
obtained in chiral perturbation theory.

\begin{figure}[tbp]
\centerline{
\includegraphics[width=8cm]{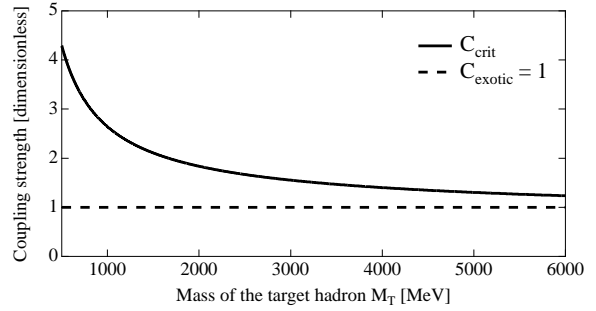}
}
\caption{Critical coupling strength $C_{\text{crit}}$ for $f=93$ MeV 
and $m=368$ MeV (Solid line). The dashed line denotes the universal 
attractive coupling strength in exotic channels $C_{\rm exotic}=1$.}
\label{fig:critical}
\end{figure}

With the renormalization condition~\eqref{eq:regucond}, we derive the 
smallest attractive coupling strength $C_{\rm crit}$ to produce a bound
state of the NG boson with the target hadron as
\begin{align}
    C_{\text{crit}}= \frac{2f^2 }{m\bigl[-G(M_{T}+m)\bigr]} .
    \label{eq:WTcritical}
\end{align}
Using $f=93$ MeV and $m=368$ MeV, we plot $C_{\text{crit}}$ in 
Fig.~\ref{fig:critical} as a function of $M_{T}$ together with the universal 
attractive strength $C_{\rm exotic}=1$ found in 
Eq.~\eqref{eq:Exoticattraction}. It is seen that $C_{\rm exotic}$ is not 
strong enough to generate a bound state in this region. We also examine 
the dependence of the $C_{\text{crit}}$ on $f$ and $m$ and find that
$C_{\rm crit}$ becomes smaller, either for the larger $m$ or the smaller $f$.
However, $C_{\rm crit}$ cannot be smaller than $C_{\rm exotic}=1$ in 
physically reasonable parameter region.

%--figure---------------------------------
%--figure---------------------------------

% large Nc

For a baryon target, we can evaluate the WT interaction in large-$N_c$ 
limit using the analytic form of the coupling strength~\eqref{eq:WTintfinal}.
With the standard extension of the flavor representation $[p,q+(N_c-3)/2]$ 
with fixed baryon spin~\cite{Dashen:1993jt}, we derive the nontrivial linear 
$N_c$ dependence of the coupling strength $C_{\alpha ,T}(N_c)$, as shown in 
Table~\ref{tbl:ECtable}. We find that the interaction in exotic channels has 
negative dependence of $N_c$. Combining with the $1/N_c$ dependence of the 
$1/f^2$ factor, we show that no attractive interaction exists in the 
large-$N_c$ limit.

% summary,

We have studied the exotic hadrons in $s$-wave chiral dynamics. Based on the 
group theoretical argument and the general principle of the scattering 
theory, we have drawn the following two conclusions:
\begin{itemize}
    \item  Attractive interaction in exotic channels is 
    universal $C_{\rm exotic}=1$. \\

    \item  The interaction is smaller than the critical 
    value to generate a bound state $C_{\rm exotic}<C_{\rm crit}$.

\end{itemize}
Combining these two, we have shown that the exotic hadrons are not 
generated in $s$-wave scattering of the NG boson with a target hadron 
in the SU(3) symmetric limit.

It should be, however, noted that the present analysis is based on the 
simplest framework with the leading order interaction in the SU(3) limit.
In practice, the effect of the flavor SU(3) breaking and higher order 
terms in the chiral Lagrangian should be taken into 
account~\cite{Hyodo:2003qa,Borasoy:2004kk,Borasoy:2005ie,Oller:2005ig,Oller:2006jw} .
In addition, we do not exclude the existence of the exotic state with 
different origins, such as genuine quark states or the rotational excitation 
of the chiral solitons. Nonetheless, the conclusion drawn here should 
contribute to partly explain the difficulty to observe of the exotic hadrons
in Nature.

The authors are grateful to Prof.\ M.\ Oka for helpful discussion. We also 
thank Professor V.\ Kopeliovich for useful comments on exoticness. T.~H. 
thanks the Japan Society for the Promotion of Science (JSPS) for financial 
support.  This work is supported in part by the Grant for Scientific 
Research (No.\ 17959600, No.\ 18042001, and No.\ 16540252) and by 
Grant-in-Aid for the 21st Century COE "Center for Diversity and Universality
in Physics" from the Ministry of Education, Culture, Sports, Science and 
Technology (MEXT) of Japan.

% \bibliographystyle{h-physrev3}
% \bibliography{refs,refs05,refs00,myrefs}

%
% BibTeX users please use
% \bibliographystyle{}
% \bibliography{}
%
% Non-BibTeX users please use
% % \begin{thebibliography}{}
% % %
% % % and use \bibitem to create references.
% % %
% % \bibitem{RefJ}
% % % Format for Journal Reference
% % Author, Journal \textbf{Volume}, (year) page numbers.
% % % Format for books
% % \bibitem{RefB}
% % Author, \textit{Book title} (Publisher, place year) page numbers
% % % etc
% % \end{thebibliography}

\end{document}